\documentclass[twocolumn]{svjour3}

\smartqed  
\usepackage{graphicx}

\usepackage[caption=false,font=footnotesize]{subfig}

\usepackage{xspace}

\usepackage[utf8]{inputenc}

\usepackage{amssymb}
\usepackage{amsmath}
\usepackage{algorithm}
\usepackage{algpseudocode}
\usepackage{enumitem}  
\usepackage[hidelinks]{hyperref}
\usepackage{tikz}

\usetikzlibrary{automata,positioning,arrows}
\usetikzlibrary{calc}
\tikzset{initial text={}}

\newcommand{\fett}[1]{\boldsymbol #1}
\newcommand{\albet}{\mathcal A}

\newcommand{\citlab}{CITlab\xspace}

\DeclareMathOperator*{\F}{\mathcal F}
\DeclareMathOperator*{\p}{P}

\DeclareMathOperator*{\argmax}{arg max}

\begin{document}
%
\title{System Description of CITlab's Recognition \& Retrieval Engine for ICDAR2017 Competition on Information Extraction in Historical Handwritten Records}


\author{
Tobias Strau\ss \and Max Weidemann \and Johannes Michael \and Gundram Leifert \and Tobias Gr\"uning \and Roger Labahn
} 

\institute{Tobias Strau\ss \and Max Weidemann \and Johannes Michael \and Gundram Leifert  \and Roger Labahn \at
              Institute of Mathematics \\
              University of Rostock\\
              18051 Rostock\\
              Tel.: +49-381-4986633\\
              \email{\{tobias.strauss, max.weidemann, johannes.michael, gundram.leifert,  roger.labahn\}@uni-rostock.de}           
            \and
            Tobias Gr\"uning \at
            Planet AI\\
            Warnowufer 60\\ 18057 Rostock\\
            \email{tobias.gruening@planet.de}
}

\date{Received: date / Accepted: date}


\maketitle

\begin{abstract}
We present a recognition and retrieval system for the ICDAR2017 Competition on Information Extraction in Historical Handwritten Records which successfully infers person names and other data from marriage records. The system extracts information from the line images with a high accuracy and outperforms the baseline. The optical model is based on Neural Networks. To infer the desired information, regular expressions are used to describe the set of feasible words sequences. 
\end{abstract}


\section{Introduction}
There is a huge amount of handwritten texts containing information of past times which are valuable but not yet accessible. The ICDAR2017 Competition on Information Extraction in Historical Handwritten Records encourages research in the field of automatic retrieval systems by providing training data from marriage records. 

We present a bottom-up approach which processes the writing resulting in a matrix of probabilities per character and position. A two step process finds the most likely character sequence according to this matrix and previously defined regular expressions covering the expected structure and assigns information containing parts of this sequence to the specific categories. 


\section{Task}
\begin{figure*}[t]
\centering
\subfloat[dit dia rebere de Luys (name/H)  Burgues (surname/H) llibrater (occupation/H) de Bara (location/H) fill de Jua (name/H's father)
]{\includegraphics[width=0.95\textwidth]{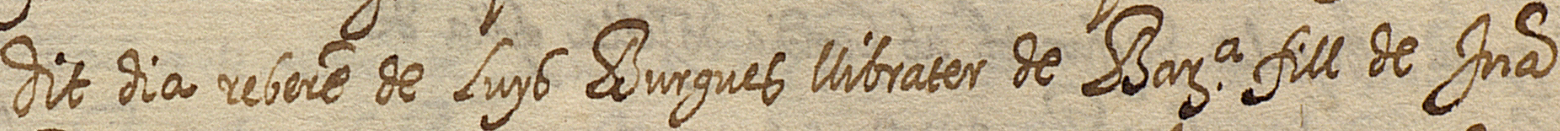}}
\hfil
\subfloat[Burgues (surname/H's father) llibrater (occupation/H's father) y de Angela (name/H's mother) defuncts ab Anna (name/W) viuda (state/W) de]{\includegraphics[width=0.95\textwidth]{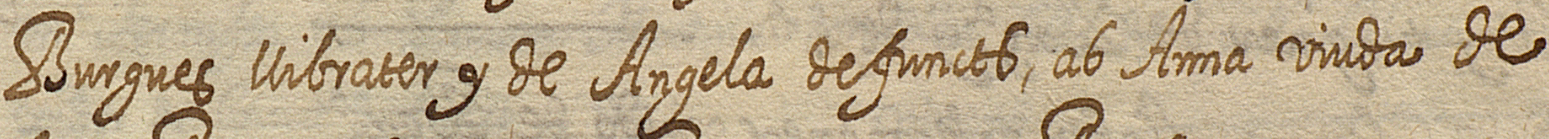}}
\hfil
\subfloat[Jua (name/other person) Basili (surname/other person) sastre (occupation/wife) de Bara (location/wife) mori en Bara]{\includegraphics[width=0.95\textwidth]{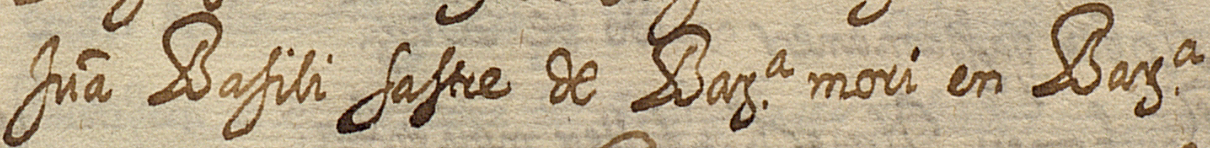}}
\caption{Sample record from the Esposalles data set. Categories and corresponding person class for words of interest in parenthesis. H and W mean husband and wife, respectively.}\label{fig:sample}
\end{figure*}

The data set consists of well-written marriage records of the 17th century from the Esposalles database. The task is to extract words of categories of interest like name, surname, location and state (Track 1) and assign them to persons like husband, wife, husband's father, wife's mother etc. (Track 2) from the given line images. A sample record is given in Fig. \ref{fig:sample}. 

The organizers provided 970 records (consisting of 3070 lines) for training and validation including transcriptions, categories and person classes. The test set comprises 757 lines from 253 records. 
The major problem with the data set is to parse the variations of the language. Promising sequence 2 sequence approaches (see \cite{sutskever2014sequence}) could solve this issue in the future without manual effort which is still necessary for the proposed system.

\section{Recognition Engine and Retrieval}
\subsection{Preprocessing}
Given the line polygon, we apply certain standard preprocessing routines, i.e.
\begin{itemize}
	\item image normalization: contrast enhancement (no binarization), size;
	\item writing normalization: line bends, line skew, script slant.
\end{itemize}
Then, images are further unified by \citlab's proprietary writing normalization:  The writing's main body is placed in the center part of an image of fixed 96px height. While the length-height ratio of the main body stays untouched, the  ascenders and descenders are squashed to focus the network's attention on the more informative main body.
\subsection{Neural Network}
The preprocessed images are fed into a neural network of the architecture described in Table \ref{tab:network}. The implementation is based on TensorFlow (see \cite{abadi2016tensorflow}). The three convolutional layers additionally apply  batch normalization (see \cite{ioffe2015batch}) before and local response normalization (see \cite{krizhevsky2012imagenet}) after applying the ReLU activation function. The BLSTM layers are trained with dropout (applied to the output and keep ratio of 0.5, see \cite{gal2016theoretically}).

\setlength{\tabcolsep}{5pt}
\begin{table}[h]
	\centering
\caption{Network layer from input (left) to output (right)}\label{tab:network}
\begin{tabular}{lcccccc}
	\hline
	&conv&conv&BLSTM&conv&BLSTM&fully\\\hline
 Neurons	& 8 &  32 & 256 & 64 & 512 & 62 \\
 stride & 4x3& 4x3&&1x2
 \\\hline
\end{tabular}
\end{table}

The last layer is the fully-connected layer and contains 62 neurons. One of these neurons represents a garbage label $\oslash$ (not-a-character or NaC in the following) and the others correspond to the 61 characters appearing in the ground truth. We denote the \emph{character set} of the 61 characters by $\albet$ and \emph{label set}  $\albet \cup \{\oslash\}$ by $\albet'$ here and after. The loss function is the typical CTC-loss (see \cite{graves2006connectionist}). The network is trained 150 epochs by RMSProp  (see \cite{tieleman2012lecture}) where one epoch contains 4096 randomly sampled line images.  The initial learning rate is 0.002 and decayed after every third epoch by a factor of 0.95.

The output of the last layer is softmax transformed such that the output of the neural network is a matrix $\fett Y\in[0,1]^{T\times 62}$ of variable length $T$. For each row $t$, $\sum_{l\in\albet'} y_{t,l}=1$. We call $\fett Y$ \emph{ConfMat}.

\subsection{Decoding}
Certain lines (and thus the corresponding  ConfMats) belong to the same record. These ConfMats are concatenated to one whole ConfMat per record. The encoded text follows specific rules which can be formulated as regular expressions. To decode the most likely character sequence according to a regular expression, we use the method described in \cite{strauss2016decoding}.

Let $\F: \albet' \rightarrow \albet$ be the mapping which deletes consecutive identical labels and removes all NaCs, e.g. $\F(\oslash a \oslash a b)= \F(\oslash a \oslash aaa b) = aab$.  The probability of a label sequence $\fett l$ given a line image $\fett X$ is calculated by $\p(\fett l \mid \fett X) = \prod_{t=1}^{T} y_{t,l_t}$ if the ConfMat and the label sequence are both of length $T$ and $0$ otherwise. 
The most likely character sequence $\fett z$ maximizes $\sum_{\fett l \in \F^{-1}(\fett z)}\p(\fett l \mid \fett X)$. Since there is typically one dominant label sequence, we substitute the sum by the maximum: 
$$\fett z^*:= \argmax_{\fett z} \max_{\fett l \in \F^{-1}(\fett z)}\p\nolimits(\fett l \mid \fett X)$$

The proposed method is based on two steps: A first coarse labeling is done by a regular expression which splits the whole record ConfMat into regions corresponding to the various persons: husband, wife and their parents. The regular expression is generated manually and includes none of the given vocabularies. The structure of the expression is simple: the regions are identified by several keywords which are followed by a region corresponding to a specific person. 

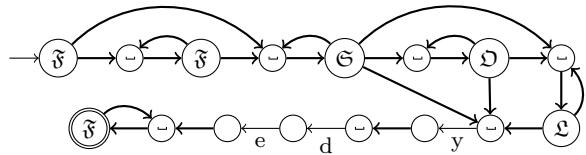
\begin{figure}
\tikzstyle{every state}=[inner sep=2pt,minimum size=10pt]

\begin{tikzpicture}[align=center,node distance=0.5cm] 
	\node[state,initial] (f1) at (0,0) {$\mathfrak F$}; 
	\node[state, right= of f1] (s1) {\textvisiblespace};
	\node[state, right= of s1] (f2) {$\mathfrak{F}$};
	\node[state, right= of f2] (s2) {\textvisiblespace};
	\node[state, right= of s2] (S) {$\mathfrak{S}$};
	\node[state, right= of S] (s3) {\textvisiblespace};
	\node[state, right= of s3] (O) {$\mathfrak{O}$};
	\node[state, right= of O] (s4) {\textvisiblespace};
	\node[state, below= of s4] (L) {$\mathfrak{L}$};
	
	\node[state, left= of L] (s5) {\textvisiblespace};
	\node[state, left= of s5] (y) {};
	\node[state, left= of y] (s6) {\textvisiblespace};
	\node[state, left= of s6] (d) {};
	\node[state, left= of d] (e) {};
	\node[state, left= of e] (s7) {\textvisiblespace};
	\node[state, left= of s7,accepting] (f3) {$\mathfrak{F}$};
	
	\draw[out=45,in=-45, ->,thick] (L) to (s4);
	\draw[out=45,in=135, ->,thick] (f3) to (s7);
	\foreach \source/\target in {S/s2, f2/s1, O/s3}
		\draw[out=135,in=45, ->,thick] (\source) to (\target);
	\foreach \source/\target in {f1/s1, s1/f2, f2/s2, s2/S, S/s3, s3/O, O/s4, s4/L, L/s5, O/s5, S/s5, y/s6, e/s7, s7/f3}
		\draw[->,thick] (\source) -- (\target);
	
	\foreach \source/\target in {f1/s2, S/s4}
		\draw[out=45,in=135, ->,thick] (\source) to (\target);
		
	\draw[->] (s5) -- node[below]{y} (y);	
	\draw[->] (s6) -- node[below]{d} (d);
	\draw[->] (d) -- node[below]{e} (e);
	 
\end{tikzpicture}
\caption{Simplified automaton accepting the information of the parents. Nodes with Letters or symbols inside symbolize subautomata of dictionaries. After one or more first names ($\mathfrak{F}$), at least one surname ($\mathfrak{S}$) has to be recognized followed by optional occupations ($\mathfrak{O}$) and locations ($\mathfrak L$). The \textvisiblespace{} automaton accepts concatenations of spaces and linebreaks. Thick arrows represent multi arcs involving at least one dictionary subautomaton.}\label{fig:automaton}
\end{figure}

The second step processes these regions corresponding to a specific person separately (see Figure \ref{fig:automaton}). Here, the task is to identify names, locations etc. Incorporating a vocabulary yields more reliable transcriptions than using the most likely network output directly. Thus, we include the provided vocabularies into the regular expression. Only the general category vocabularies are used ignoring e.g. those corresponding to specific persons. Even the surname vocabulary alone comprises more than 1200 names such that a beam search is required to decode the most likely character sequence.

The neural network does not model the prior probability $\p(\fett z)$ of a word $\fett z$ correctly.  A simple application of Bayes law (see \cite{strauss2016decoding}) yields a corrected probability $\p_{\mathcal T}(\fett z \mid \fett X)$ of the character sequence $\fett z$ given the image $\fett X$

$$
\p\nolimits_{\mathcal T}(\fett z \mid \fett X)\propto\frac{\p_{\mathcal T}(\fett z)}{\p_{\mathcal S}(\fett z)}\p\nolimits_{\mathcal S}(\fett z \mid \fett X)	
$$
up to a normalization which is the same for any character sequence given the same image $\fett X$. Here, $\p_{\mathcal S}(\fett z\mid \fett X)$ represent the probability of the neural network as defined above. $\p_{\mathcal S}(\fett z)$ is the prior probability implicitly learned by the neural network. This term cannot be measured directly and has to be estimated. The term $\p_{\mathcal T}(\fett z)$ is the true (or at least better) prior probability of the character sequence $\fett z$. 

In the competition, the decoded character sequence maximizes
$$
\p\nolimits_{\mathcal T}(\fett z)\p\nolimits_{\mathcal S}(\fett z \mid \fett X).
$$
That means, $\p_{\mathcal S}(\fett z)$ is assumed to be uniformly distributed over $\albet^T$ (which is not true).  For $\fett z= (\fett z_{1},\dots,\fett z_n)$, $\p_{\mathcal T}(\fett z)$ is approximated by the product of the relative frequencies of its subwords from the corresponding vocabularies, i.e., $\prod_{i}\p_{\mathcal T}(\fett z_i)$ (if we ignore spaces and words that are not from vocabularies). Any conditional dependency (e.g. the probability of a location or occupation after the surname) is ignored.

\setlength{\tabcolsep}{3pt}
\begin{table*}[h]
	\centering
	\caption{Competiotion score (based on CER) for CITlab's recognition and retrieval system on the track complete.} \label{tab:results}
	\begin{tabular}{*{14}{c}}
		\hline
		\multicolumn{5}{c}{husband}  & \multicolumn{4}{|c|}{husband's father} &  \multicolumn{2}{c}{husband's mother} &  \multicolumn{3}{|c}{other person}\\
		\hline
		name & surname & state & location & occupation & \multicolumn{1}{|l}{name} & surname &  location & occupation  & \multicolumn{1}{|l}{name} & surname& \multicolumn{1}{|l}{name} & surname & state\\
		\hline
		96.10&88.85&92.42&90.42 & 88.49 &
		94.28& 86.57&  78.61 & 92.07 &
		96.17 & 0 &
		93.93 &  88.06 & 0\\\hline\\
		\cline{1-11}
		\multicolumn{5}{c}{wife} & \multicolumn{4}{|c|}{wife's father} &  \multicolumn{2}{c}{wife's mother} \\
		name & surname & state & location & occupation & \multicolumn{1}{|l}{name} & surname & location & occupation  & \multicolumn{1}{|l}{name} & surname \\
		\cline{1-11}
		98.49 & 36.57 & 97.13 &66.73 &  91.43  & 
		94.42 & 87.43 & 89.29 & 89.17 & 
		95.90 & - & 
		\\\cline{1-11}
	\end{tabular}
\end{table*}

To allow also out-of-vocabulary words, we added the most likely characters per position instead of first name or surname. The prior for such an out-of-vocabulary word is a combination of a character probability and a word probability which is negligible small compared ot the relative frequency of any vocabulary word. 

\section{Competition results}

We briefly report the results of the complete track. Details can be found in \cite{fornes2017icdar2017}.  The score of the ICDAR2017 Competition on Information Extraction in Historical Handwritten Records is equal to the character accuracy if the category and person (basic track: only category) are correct and 0 otherwise.

Besides the baseline and our systems, there is no other submission at line level. Another track of the same competition provides a word segmentation instead of the line as whole image. Task and score are the same for both levels. The best retrieval system at word level performs slightly better than our best system (overall score of 91.97 against 91.56).

%
 
\subsection*{Discussion}
In Table \ref{tab:results}, the competition results are presented (as given in the article of the organizers \cite{fornes2017icdar2017}). We find systematical gaps e.g. the recognition of the name of any person is always more reliable than the surname. The organizers explained this by the greater variability of surnames.  

In total the scale of the results are similar except for the categories husband's mother's name, other person's state, wife's surname and wife's location. The first two categories are not considered by our expression and also the other competition participants returned 0 scores. This indicates that these categories are rarely presented in the training data and validation data. For the latter two categories the regular expression seems to fit not very well.

\section{Conclusion and Outlook}
\label{sec:conc}
We presented a retrieval algorithm for the ICDAR2017 Competition on Information Extraction in Historical Handwritten Records.  The task is to extract information of the various persons from the lines. The proposed system is based on deep recurrent neural networks. Regular expressions are defined to decode the output. The system is able to infer most of the categories with high precision. 

A drawback of the proposed system is the relatively high manual effort to define the precise regular expression. In the future, we will work on reducing this effort either by learning the regular expression automatically or applying the powerful seq2seq models which have shown to cope with such kind of tasks.

\section*{Acknowledgment}
	This work was partially funded by the European Union's Horizon 2020 research and innovation programme under grant agreement No 674943 (READ -- Recognition and Enrichment of Archival Documents).

We gratefully acknowledge the support of NVIDIA Corporation with the donation of the Titan X Pascal GPU used for this research.

\bibliographystyle{apalike}
\bibliography{lit.bib}

%




%


\end{document}